\documentclass[%
 reprint,
 amsmath,amssymb,
 aps,
 prx,
floatfix,
]{revtex4-2}

\usepackage[mode=buildnew]{standalone}
\usepackage{graphicx}
\usepackage{dcolumn}
\usepackage{bm}
\usepackage{braket}
\usepackage{xcolor}
\usepackage[utf8]{inputenc}
\usepackage[T1]{fontenc}
\usepackage[english]{babel}
\usepackage{letltxmacro}
\usepackage{mathtools}
\usepackage{ifthen}
\usepackage{ragged2e}
\usepackage{placeins}
\usepackage{enumerate}
\usepackage{siunitx}

\usepackage{tikz-cd}
\usepackage{tikz}
\usetikzlibrary{automata,arrows,positioning,calc,fit,arrows.meta,shapes}

\LetLtxMacro{\ORIGselectlanguage}{\selectlanguage}
\makeatletter
\DeclareRobustCommand{\selectlanguage}[1]{%
  \@ifundefined{alias@\string#1}
    {\ORIGselectlanguage{#1}}
    {\begingroup\edef\x{\endgroup
       \noexpand\ORIGselectlanguage{\@nameuse{alias@#1}}}\x}%
}
\newcommand{\definelanguagealias}[2]{%
  \@namedef{alias@#1}{#2}%
}

\definelanguagealias{en}{english}

\newcommand\numberthis{\addtocounter{equation}{1}\tag{\theequation}}
\newcommand{\Path}[1]{\mathcal{P} \ifthenelse{\equal{#1}{}}{}{[#1]}}                    
\newcommand{\Pathr}[1]{\widetilde{\mathcal{P}}\ifthenelse{\equal{#1}{}}{}{[#1]}}        
\newcommand{\mean}[1]{\left\langle #1 \right\rangle}                                    
\newcommand{\obs}{\mathcal{O}}                                                          
\newcommand{\mev}[1]{\mathcal{#1}}                                                      
\newcommand{\ab}{\tau}                                                                  
\newcommand{\rev}[1]{\widetilde{#1}}                                                    
\newcommand{\wtd}{\psi}                                                                 
\newcommand{\wtdto}[2]{\wtd_{#1\to #2}}                                                 
\renewcommand{\d}{\text{d}}

\newcommand{\traj}{\Gamma}                                                              
\newcommand{\trajr}{\rev{\traj}}                                                        
\newcommand{\Prob}[1]{P(#1)}                                                            
\newcommand{\snip}{\Gamma^\text{s}}                                                     
\newcommand{\rsnip}{\rev{\Gamma}^\text{s}}                                              
\newcommand{\Pathb}[1]{\mathcal{P}^\text{cf} \ifthenelse{\equal{#1}{}}{}{[#1]}}          
\newcommand{\aff}{\mathcal{A}}                                                          
\newcommand{\cyc}{\mathcal{C}}                                                          
\newcommand{\cg}{\mathcal{C}}                                                           
\newcommand{\tr}{\mathcal{T}}                                                           
\newcommand{\R}{\mathcal{R}}                                                            
\newcommand{\A}{A}                                                                      
\newcommand{\fric}{\gamma}                                                                   

\begin{document}
\preprint{APS/123-QED}

\title{Fluctuating Entropy Production on the Coarse-Grained Level: \\Inference and Localization of Irreversibility}

\author{Julius Degünther}
\thanks{J.D. and J.v.d.M. contributed equally to this work.}
\author{Jann van der Meer}
\thanks{J.D. and J.v.d.M. contributed equally to this work.}
\author{Udo Seifert}
\affiliation{
 II. Institut für Theoretische Physik, Universität Stuttgart, 70550 Stuttgart, Germany
}

\date{\today}

\begin{abstract}
Stochastic thermodynamics provides the framework to analyze thermodynamic laws and quantities along individual trajectories of small but fully observable systems. If the observable level fails to capture all relevant degrees of freedom, some form of effective, coarse-grained dynamics naturally emerges for which the principles of stochastic thermodynamics generally cease to be applicable straightforwardly. Our work unifies the notion of entropy production along an individual trajectory with that of a coarse-grained dynamics by establishing a framework based on snippets and Markovian events as fundamental building blocks. A key asset of a trajectory-based fluctuating entropy production is the ability to localize individual contributions to the total entropy production in time and space. As an illustration and potential application for inference we introduce a method for the detection of hidden driving. The framework applies equally to even and odd variables and, therefore, includes the peculiar case of entropy production in underdamped Langevin dynamics. 
\end{abstract}

\maketitle


\section{Introduction}

How can we apply the principles of stochastic thermodynamics to effective descriptions? Originally, stochastic thermodynamics emerged as a framework to identify and formulate thermodynamic laws for small-scale systems coupled to the environment via, \textit{e.g.}, random thermal fluctuations and driving forces \cite{seif12, vdb15}. As the dynamics itself and, consequently, associated thermodynamic quantities like heat, work or entropy become inherently stochastic, the notion of a 'system' only makes sense if there is a meaningful, 'clean' separation from the environment. More specifically, a meaningful notion of energy or entropy requires a notion of 'thermodynamic consistency' when implementing external effects into the stochastic dynamics of the system.

Thus, if such a system can be identified, \textit{e.g.}, in the form of a Langevin equation or Markovian dynamics on a discrete set of states, the framework of stochastic thermodynamics provides far-reaching, universal relations like, \textit{e.g.}, the Jarzynski equality and its generalizations \cite{jarz97, croo98, jarz11}, different formulations of fluctuation theorems \cite{kurc98, lebo99, croo99, croo00, espo10b}, the Hatano-Sasa relation \cite{hata01} and the Harada-Sasa relation \cite{hara05}, many of which have been realised experimentally, as reviewed in \cite{cili17}.

These results are formulated for a complete system in the sense that the stochastic dynamics gives rise to trajectories that include all relevant degrees of freedom. With complex real-world systems in mind, we can ask what remains if we are unable to observe the full system or, even worse, if a clear distinction between system and environment cannot be made based on the available data. Such questions have seen growing interest in recent research, leading to a point where the study of partially accessible information might be regarded as an emerging pillar of stochastic thermodynamics in its own right.

Up to now, the study of stochastic thermodynamics of partial information contains two mostly disjoined aspects. Existing methods of thermodynamic inference \cite{seif19} mainly focus on extracting or deducing particular averaged quantities based on thermodynamic relations and limited access to observables. However, a partially accessible system features its own effective dynamical laws, which emerge as a projection of the underlying dynamics. Thus, inference techniques formulated on such a dynamical level should be able to provide thermodynamic bounds on a fluctuating level beyond simple averages.

In contrast, thermodynamic inference employs a variety of methods primarily aiming at the estimation of mean entropy production. Most prominently, various formulations and generalizations of the thermodynamic uncertainty relation \cite{bara15, ging16, horo20} provide bounds on the minimal thermodynamic cost to achieve a certain precision of, \textit{e.g.}, a current. Furthermore, lower bounds on entropy production can also be based on an identification as a Kullback-Leibler divergence \cite{kawa07, rold10, bisk17, mart19, vdm22b, haru22, kapu22}, the speed at which the system evolves in time \cite{shir18, ito20}, stopping times \cite{neri17}, waiting times \cite{skin21} and counting events \cite{piet23}. Novel recent approaches include dynamical correlations into entropy estimation techniques by relating entropy production to the asymmetry of cross-correlations of accessible observables \cite{ohga23, shir23, vu23}, their power spectral density \cite{dech23a}, correlation times \cite{dech23}, or, more technically, the spectrum associated with the dynamics \cite{kolc23}. Beyond estimating the average value of entropy production, inference techniques for the topology of the underlying system \cite{li13, thor20} and its driving affinities \cite{vdm22, ohga23} have also been proposed. 

The dynamics of coarse-grained systems received much attention in earlier works on stochastic thermodynamics, usually in the context of state lumping \cite{espo12, seif20a}. More recently, a gradual paradigm shift in the conception of coarse graining challenges established paradigms \cite{hart21b, gode23}. To fully understand real-world scenarios it seems indispensable to find descriptions for dynamics on the coarse-grained level, with recent approaches including milestoning \cite{elbe20, hart21a} or semi-Markovian dynamics \cite{vdm22}. 

The present work unifies the notion of entropy production along a single trajectory with that of coarse-grained dynamics. We establish a framework that allows us to identify entropy production along individual trajectories that retains meaning on the levels of both the underlying system and the coarse-grained, observable one, even if the coarse-grained dynamics does not obey simple rules like a master or Langevin equation. The concept of fluctuating entropy production for individual, coarse-grained paths comes along with the ability to localize individual contributions to the total entropy production in time and space within the coarse-grained description. This new aspect enables the inference of more detailed information far beyond the average total entropy production of a system. From a practical viewpoint, we demonstrate that the ability to attribute irreversibility to specific events or transitions between these provides access to complex settings like the localization of entropy production in the absence of any visible states as well as the qualitative and quantitative detection of irreversibility in hidden parts of the system.

The paper is structured as follows. In Section~\ref{sec:theory}, we start with an outline of our main results and the set-up under consideration. We then present our theoretical framework by introducing the two crucial concepts through which the identification of the fluctuating entropy production becomes possible. This framework enables us to to derive a method to detect hidden driving and to estimate affinities of hidden cycles, as we describe in Section~\ref{sec:affinity}. In Section~\ref{sec:justification}, we discuss the intricacies of entropy production in the presence of odd variables. In particular, we show that the present identification of entropy production is compatible with odd variables that occur in underdamped Langevin dynamics and transition-based coarse graining. We conclude in Section~\ref{sec:conclusion}.


\section{Main Results and Framework}
\label{sec:theory}

\subsection{Illustration of the main concept}
In real-world scenarios we commonly encounter systems for which a description that can be considered 'complete' or 'fundamental' in an appropriate sense would be highly complex. Thus, simplified descriptions for such systems that capture its key characteristics are indispensable. In this context, Markov networks have become one of the predominant paradigms for which the entire well-established framework of stochastic thermodynamics is available. A description based on Markov networks typically requires the presence of clear time-scale separations to separate system and environment. However, the nature of the underlying system or insufficient observational data may jeopardize such a clear time-scale separation. Our framework is designed for descriptions that lack this clear separation. 

\begin{figure}[t]
\begin{center}
    \includegraphics[scale=1]{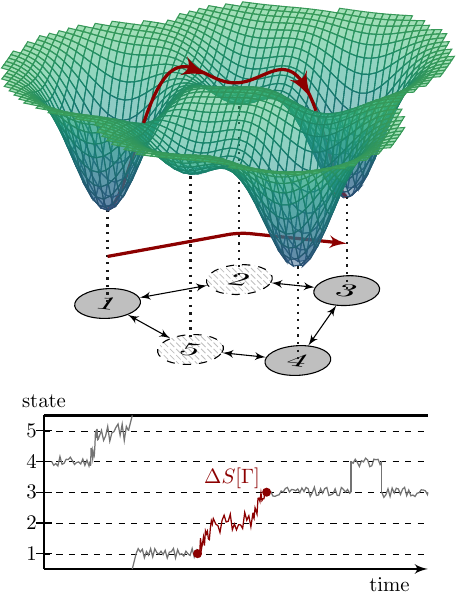}
    \caption{Projection of a continuous energy landscape onto a set of $5$ discrete states. States $2$ and $5$ cannot be modeled as Markov states due to their shallow local minima. For transitions involving these states, there is no clear time-scale separation as illustrated by the time series in the lower part. The red lines show a section of an individual trajectory. We assign an entropy production to such sections that are both localized in time and space. This identification of entropy productions retains physical significance even on the coarse-grained level.}
    \label{fig:fig0}
\end{center}
\end{figure}

These aspects are qualitatively illustrated in Figure~\ref{fig:fig0}. The energy landscape gives rise to five observable discrete states. The local minima at states $1$, $3$ and $4$ are sufficiently deep to justify a description as Markov states, in contrast to the shallow minima of states $2$ and $5$. The red line shows an exemplary trajectory, which, on the observable level, reads $1\to 2\to 3$. We will show how to identify a fluctuating entropy production for such trajectories that include non-Markovian observables like, in this case, state $2$. This coarse-grained entropy production obeys several consistency conditions and provides an estimator for the entropy production on the fundamental level. 

The ability to localize entropy production in this example translates into the ability to discern contributions to the total entropy production of, \textit{e.g.}, trajectory sections of the type $1\to 2\to 3$ from the remaining ones as indicated in the lower part of Figure~\ref{fig:fig0}. Similarly, it is possible to localize the entropy production in time. Depending on the application, one might be interested in the contributions that arise from, \textit{e.g.}, particularly fast trajectories.


\subsection{General setup}

We consider a stochastic physical system whose microscopic trajectories $\gamma$ are drawn according to some path weight $\Path{\gamma}$. For example, if the time-evolution $\gamma(t)$ obeys a Langevin or master equation, a corresponding path weight is known \cite{seif12}. We now assume that an observer can only measure particular observables, which give rise to a coarse-grained trajectory, rather than accessing this underlying level of description directly. 

The probability to observe a particular coarse-grained trajectory $\traj$ is encoded in its path weight $\Path{\traj}$ and uniquely determined by the microscopic path weight $\Path{\gamma}$ and the coarse graining $\gamma \mapsto \traj$. Similarly, the probability to observe the corresponding time-reversed trajectory $\rev{\traj}$ is determined by $\rev{\gamma}$. Thus, calculating $\Path{\rev{\traj}}$ either requires knowing the microscopic time-reversal operation $\gamma \mapsto \rev{\gamma}$ or knowing how the coarse-grained observables behave under time reversal. Reversing the coarse-grained trajectory $\traj$ generally not only means to read $\traj$ backwards, but also to modify its observables. This is the case when measuring, \textit{e.g.}, momenta or transitions. Put informally, a physical time-reversal operation not only requires 'playing the movie backwards' but also 'knowing what the movie shows', \textit{i.e.}, understanding the physical meaning of the model and the observables. This knowledge about the observables determines their time-reversal.


\subsection{Markovian events}

We define Markovian events as particular observables whose detection implies conditional independence between past and future time-evolution of $\gamma$. Such an instantaneous event determines the underlying state of the system from a dynamical point of view. We formalize the defining property for a Markovian event $\mev{I}$ as 
\begin{equation} \label{eq:mev}
    \Path{\gamma_+|\mev{I}, \gamma_-} = \Path{\gamma_+|\mev{I}}
\end{equation}
for any trajectory $\gamma_-\to\mev{I}\to\gamma_+$, which contains $\mev{I}$ between its past and future time-evolution $\gamma_-$ and $\gamma_+$, respectively. Importantly, these events are Markovian on the microscopic level.

Generally, the state of the system is fully described by including both a suitable observation $I$ and the absolute time $\ab$ at which it occurs. Therefore, a Markovian event takes the form of a tuple
\begin{equation}
    \mev{I}\equiv(I,\ab).
\end{equation} 
Suitable observations $I$ are, for example, registering a transition or the current state of the system in the case of master equation dynamics on a Markov network. We denote Markovian events by script letters $\mev{I}, \mev{J}, ...$ to emphasize an explicit time dependence. Note that in stationary systems the information about the absolute time $\ab$ is not required, which makes $I$ itself the Markovian event.

A coarse-grained trajectory may contain additional data besides Markovian events. We summarize such observables under the (potentially multidimensional) symbol $\mathcal{O}_k$, which includes any additional observables between the Markovian events $\mev{I}_{k-1}$ and $\mev{I}_k$. Thus, we write a coarse-grained trajectory as
\begin{equation} \label{eq:traj}
    \traj = \left( \mev{I}_0\xrightarrow{t_1, \obs_1} \mev{I}_1 \xrightarrow{t_2, \obs_2} \cdots \xrightarrow{t_n, \obs_n} \mev{I}_n \right)
,\end{equation}
where the waiting times $t_k$ between the Markovian events $\mev{I}_{k-1}$ and $\mev{I}_k$ are explicitly highlighted.


\subsection{Snippets}

Given a coarse-grained trajectory \eqref{eq:traj}, we define a snippet \cite{vdm22b} $\snip$ as a section of $\Gamma$, which starts with a Markovian event $\mev{I}$ and ends with the subsequent Markovian event $\mev{J}$
\begin{align} \label{eq:snip}
    \snip:\quad\quad\mev{I}&\xrightarrow[]{t,\obs}\mev{J}\\
    (I, \tau)&\xrightarrow[]{\obs}(J, \tau+t).
\end{align}
Any coarse-grained trajectory $\traj$ comprises several of these snippets. In particular, the path weight associated with the trajectory \eqref{eq:traj} factorizes into contributions from the individual snippets, 
\begin{align} \label{eq:factorization}
    \Path{\traj|\mev{I}_0}&=\Path{\snip_1|\mev{I}_0}\cdots\Path{\snip_n|\mev{I}_{n-1}},
\end{align}
where $\snip_i$ denotes the snippet between the Markovian events $\mev{I}_{i-1}$ and $\mev{I}_i$. 

In the case of stationary systems, an alternative notation for the path weight of a snippet \eqref{eq:snip} is given by
\begin{equation}
    \wtdto{I}{J}(t;\obs)\equiv\Path{\snip|\mev{I}},
\end{equation}
which concisely presents all relevant information, since the absolute time $\tau$ is not required so that the Markovian events are just the observations, \textit{i.e.}, $\mev{I}=I$ and $\mev{J}=J$. Additionally, it highlights the waiting time characteristics of the path weights $\Path{\snip|\mev{I}}$. Although useful for explicit calculations, this notation quickly appears overloaded if the system evolves in time. Therefore, we exclusively use it for systems in a stationary state. 


\subsection{Entropy production}
\label{sec:CG_entropy}

Markovian events and snippets are the fundamental building blocks that allow us to extend the concept of a fluctuating entropy production to the coarse-grained level. For a system with constant driving but not necessarily in a stationary state, we identify 
\begin{equation} \label{eq:DeltaS_snip}
    \Delta S[\snip] = \ln\frac{P(\mev{I})\Path{\snip|\mev{I}}}{P(\mev{J})\Path{\rsnip|\rev{\mev{J}}}}
\end{equation}
as the entropy production of a snippet $\snip$. It generally depends on the initial and final events $\mev{I}$ and $\mev{J}$ as well as the duration $t$ and the remaining observables $\obs$. For this identification to be physically meaningful, we assume that the entropy production on the microscopic level $\Delta s[\gamma]$ is of the form that is further detailed below in Section~\ref{sec:micro}. This assumption is justified for virtually all classes of systems that are described in the general setup and for which a physical entropy production is of interest, such as Markov networks or Langevin dynamics. For equation \eqref{eq:DeltaS_snip}, we do not need to discern between descriptions based on observables that are even under time reversal, such as state-based descriptions or overdamped Langevin dynamics and descriptions based on observables that are odd under time reversal, such as transition-based descriptions or underdamped Langevin dynamics. In Section \ref{sec:justification} we provide further details regarding the peculiar odd observables, which provide further support for the identification \eqref{eq:DeltaS_snip}.

The following properties embed this notion of entropy production along the coarse-grained trajectory \eqref{eq:DeltaS_snip} into the framework of stochastic thermodynamics. First, it is additive in the sense that the entropy production of a coarse-grained trajectory \eqref{eq:traj} is the sum of the entropy production of its snippets,
\begin{equation} \label{eq:Delta_S_additivity}
    \Delta S[\traj]=\ln\left(\frac{\Prob{\mev{I}_0}\Path{\traj|\mev{I}_0}}{\Prob{\mev{I}_n}\Path{\trajr|\rev{\mev{I}}_n}}\right)=\sum_{i=1}^n\Delta S[\snip_i],
\end{equation}
where we use equations \eqref{eq:factorization} and \eqref{eq:DeltaS_snip}. Entropy production has to be assigned in a way to avoid overcounting at the start- and endpoint, respectively. In Equation \eqref{eq:DeltaS_snip}, the entropy production of a snippet includes its initial Markovian event but not the concluding one, which becomes more apparent when rewriting the entropy production \eqref{eq:DeltaS_snip} as
\begin{equation}
    \Delta S[\snip] = \ln\frac{\Pathb{\snip|\mev{J}}}{\Path{\rsnip|\rev{\mev{J}}}}
\end{equation}
where the path weights exclude $\mev{J}$ and $\rev{\mev{J}}$ respectively. The superscript $^{\text{cf}}$ indicates that the path weight $\Pathb{\snip|\mev{J}}$ is conditioned on the final event of the trajectory snippet rather than the initial one.

\begin{figure}[t]
\begin{center}
    \includegraphics[scale=0.7]{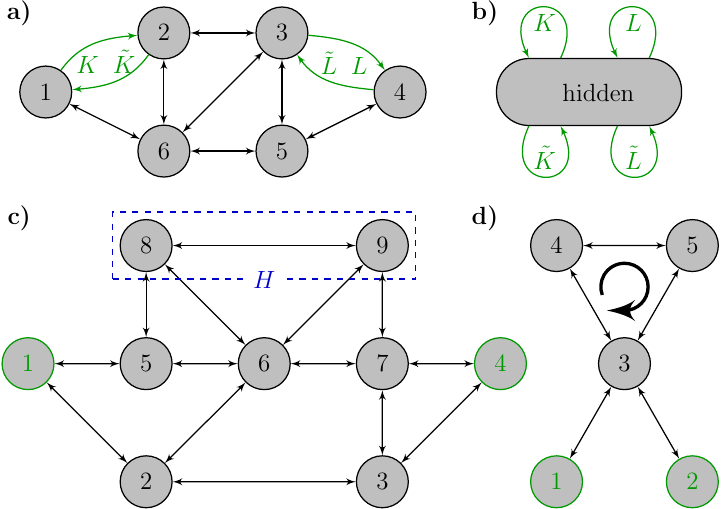}
    \caption{Markov networks and their effective descriptions. a) Markov Network with six states. We assume that only the transitions $K$ and $L$ and their reversed $\rev{K}$ and $\rev{L}$ can be observed. b) Coarse-grained description of the network shown in a). c) Markov network with nine states. We assume that only states $1$ and $4$ and the compound state $H$, consisting of states $8$ and $9$, can be observed. d) Markov Network with five states. We assume that only states $1$ and $2$ can be observed. The cycle $3\to 4\to 5\to 3$ is driven in clockwise direction. The rates used for numerical results are listed in Appendix~\ref{sec:rates}. }
    \label{fig:networks}
\end{center}
\end{figure}	

Second, the coarse-grained entropy production $\Delta S$ and the microscopic entropy production $\Delta s$ are linked by the exact relation
\begin{equation} \label{eq:s_exp_mean}
    e^{-\Delta S[\traj]} = \mean{e^{-\Delta s}|\traj},
\end{equation}
as we show in Appendix \ref{sec:proof_exp_mean}. The conditional mean $\mean{\cdot|\traj}$ denotes an average over all microscopic trajectories that are mapped to $\traj$ under coarse graining. In particular, this implies $\Delta S[\traj]=0$ if the underlying system is in equilibrium, where $\Delta s[\gamma]=0$ holds for all $\gamma$. Furthermore, equality \eqref{eq:s_exp_mean} implies the thermodynamic consistency condition 
\begin{equation} \label{eq:S_s_ineq}
    \Delta S[\traj]\leq\mean{\Delta s|\traj}
\end{equation}
for the coarse-grained entropy production. While the equality \eqref{eq:s_exp_mean} might typically be of a more theoretical rather than practical significance due to the statistically demanding nature of the mean \cite{jarz97}, the inequality \eqref{eq:S_s_ineq} will prove to be a useful and versatile tool for thermodynamic inference. In particular, it enables us to estimate entropy production localized in space and time. As a side note, the inequality \eqref{eq:S_s_ineq} directly implies the estimator
\begin{equation}
    \mean{\Delta S}\leq\mean{\Delta s} 
,\end{equation}
for the mean total entropy production of the system, which recovers the result in \cite{vdm22b} for the mean total entropy production rate following a different reasoning.

\begin{figure*}[t]
    \begin{tikzpicture}
        \node[] (test) at (0, 0) {\includegraphics[scale=0.34375]{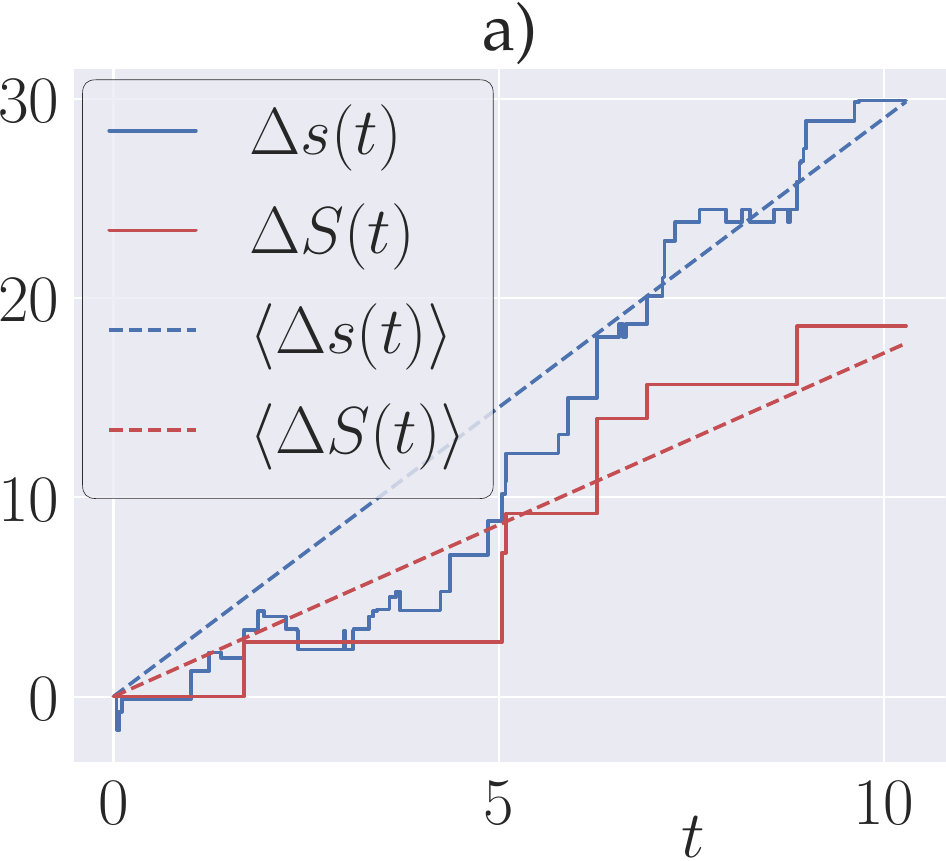}};
        \node at (6, 0) {\includegraphics[scale=0.34375]{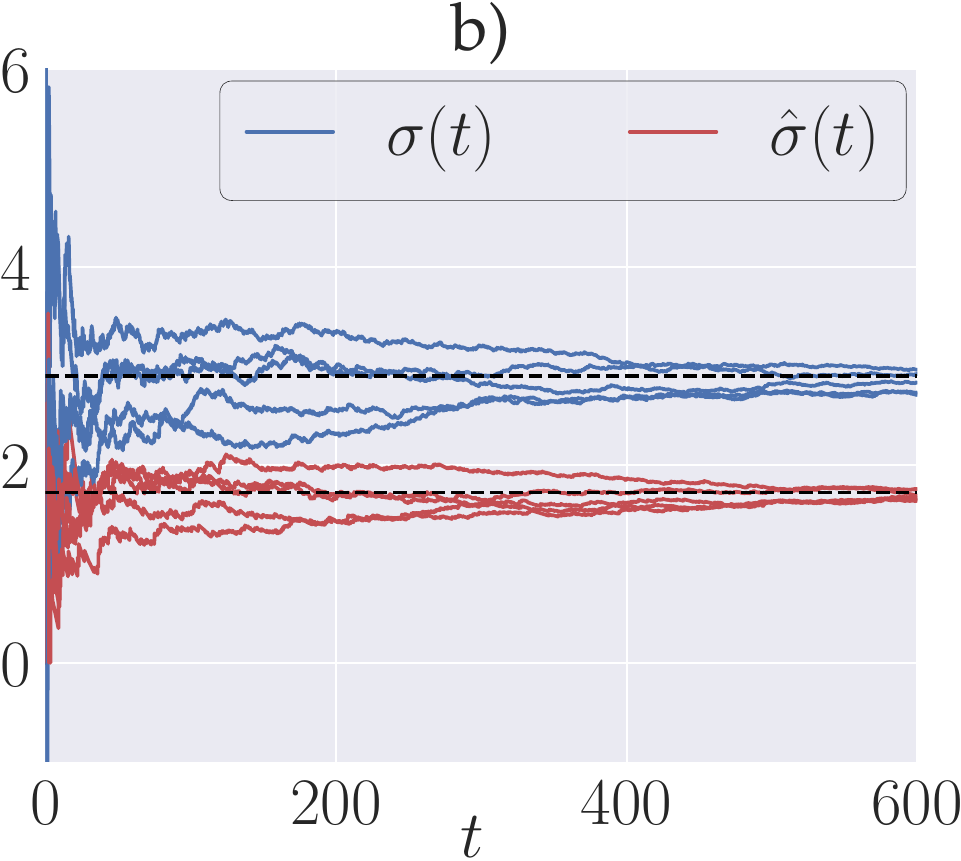}};
        \node at (11.8, 0) {\includegraphics[scale=0.34375]{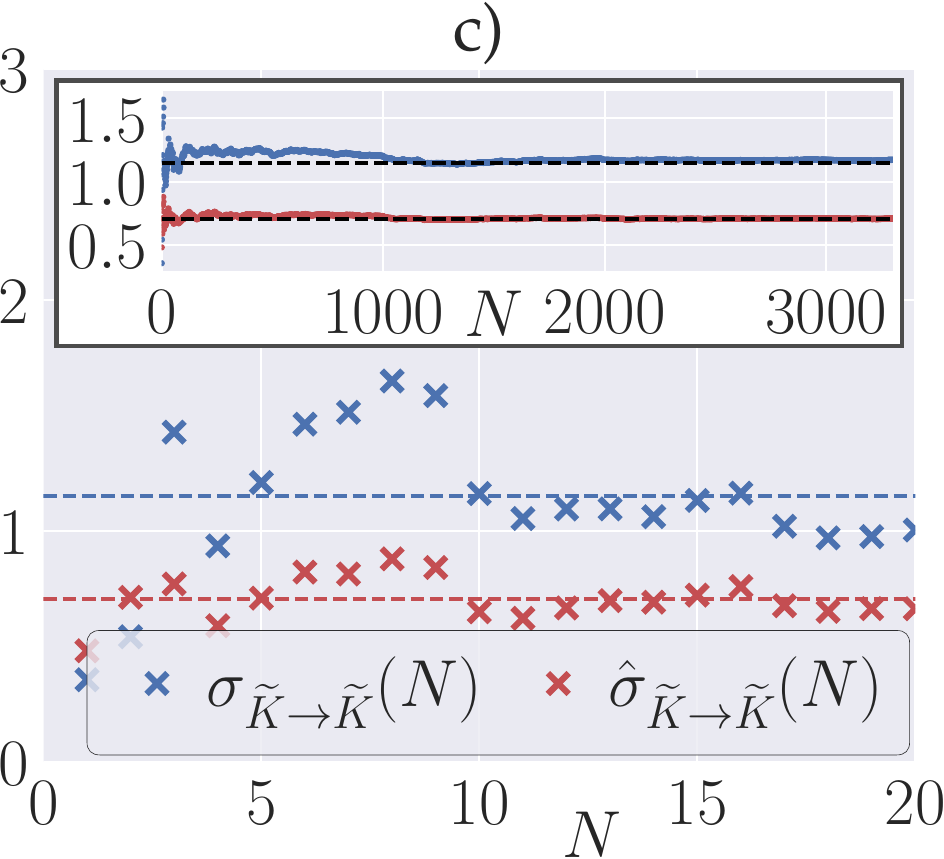}};
    \end{tikzpicture}
    \caption{Entropy production as a fluctuating quantity on the coarse-grained level. All data is generated using the network shown in Figure~\ref{fig:networks}a). Blue lines and markers indicate quantities on the microscopic level, whereas the coarse-grained level is denoted with red lines and markers. a) The solid lines show the change in total entropy as a function of time along an individual trajectory both on the microscopic and the coarse-grained level. The dashed lines show the corresponding expectation values. b) The solid lines show the total entropy production rate as a function of time along an individual trajectory for a sample of 5 trajectories on the microscopic and the coarse-grained level. The black dashed lines show the corresponding expectation values. c) The markers show the localized entropy production both on the microscopic and the coarse-grained level as defined in equations \eqref{eq:fluc_loc_sigma_micro} and \eqref{eq:fluc_loc_sigma_cg}, respectively. The statistics is obtained by considering a long trajectory in which occasionally snippets occur that start and end with the transition $\rev{K}=(21)$, see Figure~\ref{fig:networks}a). As shown by the inset, the entropy production rates $\sigma_{\rev{K}\to \rev{K}}(N)$ and $\hat{\sigma}_{\rev{K}\to \rev{K}}(N)$ converge towards their expectation values for a sufficiently high number of occurrences.
    }
    \label{fig:CG_fluc_traj.pdf}
\end{figure*}

Third, the coarse-grained entropy production $\Delta S[\traj]$ extends the notion of a fluctuating entropy production to the coarse-grained level. One of the major achievements of stochastic thermodynamics has been to identify entropy production as a fluctuating quantity along individual trajectories \cite{seif05a}, which equation \eqref{eq:DeltaS_snip} generalizes beyond microscopic trajectories. Conceptually, it is a key novelty to have a coarse-grained entropy production, which is endowed with physical meaning beyond its expectation value.


\subsection{Fluctuating entropy production on the coarse-grained level}
\label{sec:fluc_entropy}

We illustrate key aspects of the fluctuating entropy production using the concrete example shown in Figure~\ref{fig:networks}a). This microscopic system is a six-state Markov network, in which only transitions $K=(12)$ and $L=(34)$ as well as their time-reversed counterparts $\rev{K}=(21)$ and $\rev{L}=(43)$ can be observed. This gives rise to the coarse-grained description shown in Figure~\ref{fig:networks}b). For the following examples we generate the coarse-grained trajectories by simulating the microscopic ones, to which the coarse-graining is applied subsequently. This procedure is in accordance with data obtained in a real-world scenario.

The previously exclusively microscopic concept of a fluctuating entropy production can now be applied to a coarse-grained description. We define the total entropy production rate up to time $t$ as 
\begin{equation} 
\sigma(t)\equiv\Delta s(t)/t \quad \text{and} \quad
\hat{\sigma}(t)\equiv\Delta S(t)/t
\end{equation}
on the microscopic and the coarse-grained level, respectively. For some trajectories of the system from Figure~\ref{fig:networks}a), the total entropy production and the total entropy production rate along individual fluctuating trajectories on both levels is shown in Figures~\ref{fig:CG_fluc_traj.pdf}a) and b), respectively. In the coarse-grained description, the rate of jumps is lower due to the lower rate of events that contribute to the entropy production. Its expectation values fulfill inequality \eqref{eq:S_s_ineq}. The rate of coarse-grained entropy production converges towards its expectation value for each trajectory $\Gamma$ individually in the long time limit, similarly to its microscopic counterpart.

The localized entropy production exhibits a similar fluctuating behaviour. We define the entropy production rate after $N$ realizations $\{\gamma_1, ..., \gamma_N\}$ of trajectories that all start in $I$ and end in $J$ as 
\begin{equation} \label{eq:fluc_loc_sigma_micro}
    \sigma_{I\to J}(N)\equiv\frac{1}{T(N)}\sum_{i=1}^N\Delta s[\gamma_i] 
\end{equation}
and similarly its coarse-grained analogue that is defined through the corresponding snippets $\{\snip_1, ..., \snip_N\}$ as
\begin{equation} \label{eq:fluc_loc_sigma_cg}
    \hat{\sigma}_{I\to J}(N)\equiv\frac{1}{T(N)}\sum_{i=1}^N\Delta S[\snip_i] 
\end{equation}
with the accumulated duration $T(N)$ of these trajectories. This entropy production rate fluctuates with each realization as shown for $\rev{K}\to\rev{K}$ in Figure~\ref{fig:CG_fluc_traj.pdf}c). On the microscopic level, these fluctuations are due to different paths that all start with $\rev{K}$ and end with $\rev{K}$. Although these paths cannot be resolved on the coarse-grained level, different paths typically lead to a different waiting time between the initial and final event. Therefore, the fluctuating duration of the snippets leads to a fluctuating entropy production. The entropy production rate converges towards its expectation value for a sufficient number of realizations as shown by the inset in Figure~\ref{fig:CG_fluc_traj.pdf}c).


\subsection{Assumption on the microscopic level}
\label{sec:micro}

The entropy production \eqref{eq:DeltaS_snip} for the coarse-grained trajectory $\Gamma$  can be derived from a similar form on the microscopic level. We assume that the entropy production associated with a microscopic trajectory $\gamma$ that starts with $\gamma_0$ and ends with $\gamma_1$ in a possibly time-dependent but not time-dependently driven system like, \textit{e.g.}, a steady state or relaxation into it, is given by
\begin{equation} \label{eq:micro_delta_s}
    \Delta s[\gamma]=\ln\frac{P(\gamma_0)\Path{\gamma|\gamma_0}}{P(\gamma_1)\Path{\rev{\gamma}|\rev{\gamma}_0}}
.\end{equation}
Here, $\rev{\gamma}_0=\rev{\gamma_1}$, \textit{i.e.}, the time-reversed trajectory $\rev{\gamma}$ is initialized by the time-reversed final event of the original trajectory $\gamma_1$. If the process is not stationary, the probabilities $P(\gamma_0)$ and $P(\gamma_1)$ depend on the initial and final time of the process, respectively. The identification \eqref{eq:micro_delta_s} is well-known to be correct for Markovian and overdamped Langevin dynamics as well as underdamped Langevin dynamics, as further detailed in Section~\ref{sec:underdamped}. Beyond these major system classes, \eqref{eq:DeltaS_snip} cannot be derived without further knowledge about the specific system and its energetics. However, we argue that even for such systems, equation \eqref{eq:DeltaS_snip} is a reasonable starting point if appropriate Markovian events can be identified, as the identification is physically correct for all verifiable classes of systems. Note that in the case of even obserables we recover
\begin{equation} \label{eq:DFT}
     \Delta s[\gamma]=\ln\frac{\Path{\gamma}}{\Path{\rev{\gamma}}}
\end{equation}
due to $\rev{\gamma}_0=\rev{\gamma_1}=\gamma_1$. We discuss the suggested relationship between \eqref{eq:DeltaS_snip} and the detailed fluctuation theorem~\eqref{eq:DFT} later in Section~\ref{sec:discussion}.


\section{Inference of hidden driving}
\label{sec:affinity}

In a coarse-grained setup, the hidden part of a system is often not in equilibrium. Instead, it might even contain non-equilibrium processes, which do not drive any transition between two Markovian events and involve no further observables. We refer to such processes as hidden driving, which arises if, \textit{e.g.}, the hidden part of a Markov network contains cycles with non-zero affinity. Using our framework, we are generally able to detect these hidden driven cycles and to provide a lower bound on the corresponding affinities. 

In this section, we assume an underlying Markov network in a stationary state, which is necessary to have a well-defined notion of a cycle affinity of the form
\begin{equation}
    \aff_\cyc = \sum_{(ij)\in\cyc} \ln\frac{k_{ij}}{k_{ji}}
,\end{equation}
where the sum runs over all transitions $(ij)$ that point towards the same direction within a cycle $\cyc$. The rate with which a transition $(ij)$ occurs if the system is in state $i$ is denoted by $k_{ij}$. For now, we consider snippets of the form
\begin{equation}
    \snip: I\xrightarrow{t} J,
\end{equation}
and disregard any non-Markovian observables. The use of additional information $\obs$ is discussed later. We introduce the quantity
\begin{equation} \label{eq:Delta_a}
    \Delta a_{I\to J} \equiv \sup_t \Delta S[I\xrightarrow{t} J] - \inf_t \Delta S[I\xrightarrow{t} J],
\end{equation}
which allows us to bound the largest affinity of all cycles $\cyc$ hidden between $I$ and $J$ via
\begin{equation} \label{eq:affinity bound}
    \max_\cyc |\aff_\cyc| \geq\Delta a_{I\to J},
\end{equation}
as proved in Appendix~\ref{sec:affinity_proof}. This affinity bound is related to similar bounds obtained independently in a recent preprint, which are formulated for logarithmic ratios of propagators rather than waiting time distributions \cite{lian23}.

\begin{figure}[t]
\begin{center}
    \includegraphics[scale=0.7]{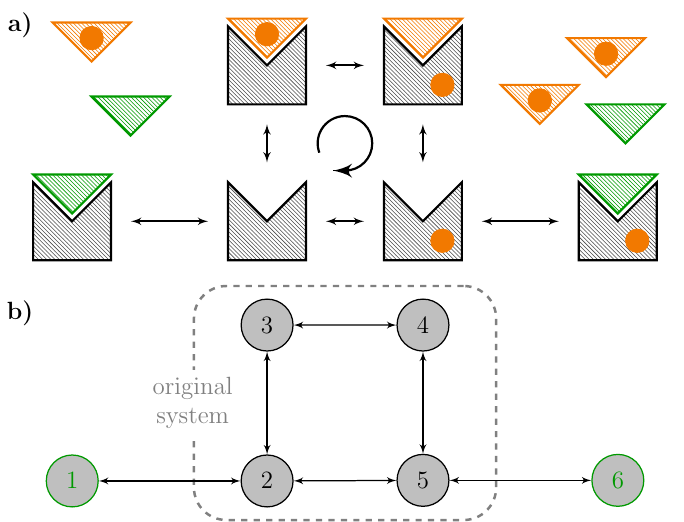}
    \caption{a) Toy model for the experimental inference of the affinity in a chemically driven system. We imagine a system in a solution of fuel carrying molecules, similar to systems driven by the hydrolysis of ATP. The original system has four states, states $2$ to $5$, which are interconnected by the following four steps: The binding and unbinding of the molecule carrying the fuel, $(23)$ and $(45)$ respectively, the consumption of the fuel $(52)$ and an intermediate step $(34)$. For all of these steps, the reverse is also possible in accordance with the condition of thermodynamic consistency, which makes the transitions bidirectional. The green triangles represent an additional chemical species, which may bind to the system if it is not occupied by a fuel-carrying molecule, \textit{i.e.}, if the system is in state $2$ or $5$. Its binding prevents any further steps along the cycle leading to the two additional states $1$ and $6$. b)~The Network corresponding to the system shown in a). We assume that only states $1$ and $6$ can be observed. The rates used for the simulations are listed in Appendix~\ref{sec:rates}.}
    \label{fig:aff_chem}
\end{center}
\end{figure}	

We demonstrate a potential biochemical application of the bound~\eqref{eq:affinity bound}, which illustrates a method to infer cycle affinities in systems that are not directly observable. As an example, we consider the toy model shown in Figure~\ref{fig:aff_chem}. We assume that we cannot measure any of the states that contribute to the cycle driving the system out of equilibrium. This remains true even after introducing an additional chemical species that is specifically designed to bind to the system in the fashion described in Figure \ref{fig:aff_chem}. However, this procedure allows to infer the chemical affinity of the driving cycle while remaining agnostic about states or transitions of the original system.

\begin{figure}[t]
\begin{center}
    \includegraphics[scale=0.34375]{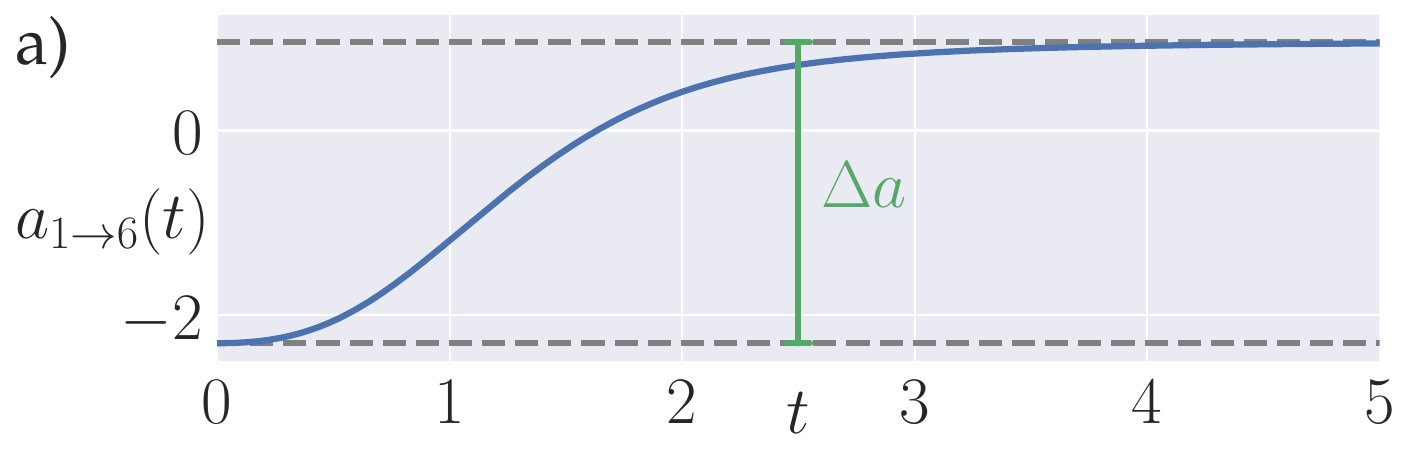}
    \includegraphics[scale=0.34375]{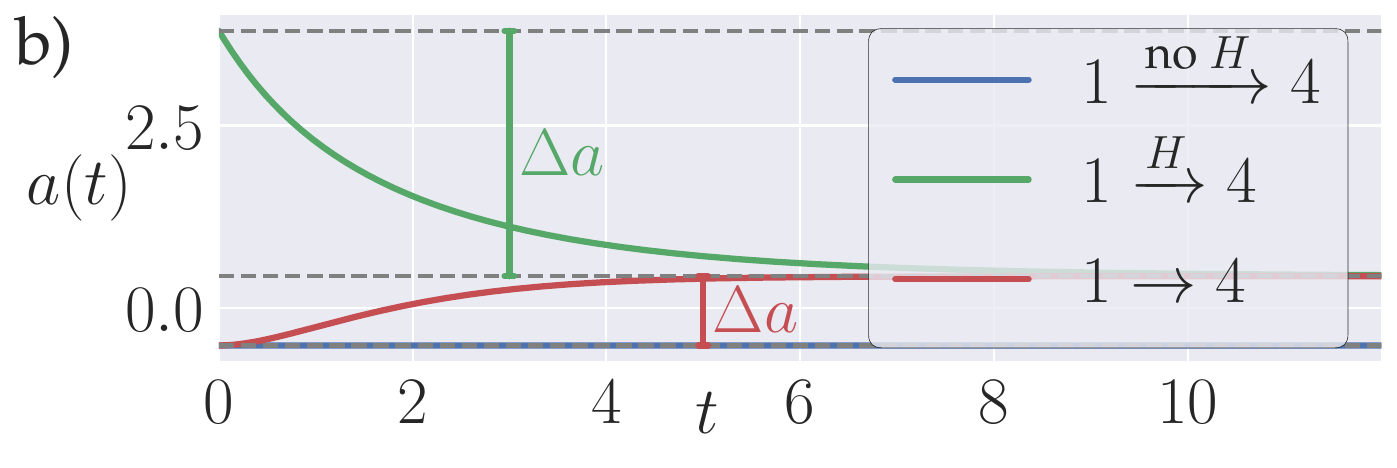}
    \caption{a) Affinity estimation for the network shown in Figure~\ref{fig:aff_chem}b) leading to $\Delta a_{1\to6}\simeq 3.3$. b) Affinity estimation for the network shown in Figure~\ref{fig:networks}c). The blue line shows $a(t)\simeq -0.51$ independent of $t$ for trajectories $1\to4$ that do not pass $H$. The green curve shows $a(t)$ for trajectories $1\to4$ that pass $H$ leading to $\Delta a\simeq 3.4$. For the red curve, the information about visiting $H$ is discarded leading to $\Delta a\simeq 0.9$.}
    \label{fig:aff_chem_plot}
\end{center}
\end{figure}

In Figure~\ref{fig:aff_chem_plot}a) we present the results of the affinity estimation for the example network shown in Figure~\ref{fig:aff_chem}. We introduce the quantity
\begin{equation}
    a_{I\to J}(t)=\ln\frac{\wtdto{I}{J}(t)}{\wtdto{\rev{J}}{\rev{I}}(t)},
\end{equation}
which is sufficient to determine $\Delta a_{I\to J}$ from equation \eqref{eq:Delta_a} through
\begin{equation} \label{eq:Delta_a_2}
    \Delta a_{I\to J} = \sup_t a_{I\to J}(t) - \inf_t a_{I\to J}(t)
\end{equation}
since the probabilities $P(I)$ and $P(J)$ do not depend on the duration of the snippet $t$. In the present case with $I=1$ and $J=6$, the estimation yields $\Delta a_{1\to 6}\simeq 3.3$ compared to the real cycle affinity of $\aff\simeq4.8$. Note that although this example relies on a unicyclic toy model, the same procedure can be applied to more complex systems, which then leads to a bound on the hidden cycle with the highest affinity. 

It is also possible to make use of additional observables besides $I$, $J$ and the time in between, but these have to satisfy an additional consistency condition. The proof of the bound $\eqref{eq:affinity bound}$ primarily relies on the construction of a suitable partial or mathematical time reversal $\R$ \cite{vdm22}, which has to be indistinguishable from the true, physical time reversal for the given coarse graining. This condition has to be met also after introducing additional observables. In our case, the involution $\R$ acts like the true time-reversal operation on some parts of the trajectory while leaving other parts unchanged as further detailed in Appendix \ref{sec:affinity_proof}. Thus, any additional observables need to be even under time reversal. In particular, we cannot use waiting times between events for data $\obs$ in general, except for the duration of the snippet $t$. 

An example for eligible further information is to include whether or not a compound state was visited during the snippet without including the time of the event. Including such additional information generally leads to an improvement of the bound \eqref{eq:affinity bound} and allows us to further localize the hidden driving. We illustrate this procedure on the example network shown in Figure~\ref{fig:networks}c) where we observe the states $1$ and $4$ as well as the lumped state $H$. If we discard the information concerning $H$, we obtain an estimate for the affinity in a similar way as for the model illustrated in Figure~\ref{fig:aff_chem},
as shown by the red line in Figure~\ref{fig:aff_chem_plot}b). When additionally observing $H$, we are able to extract two bounds. The first one applies to all cycles between $1$ and $4$ that do contain $H$, shown by the green line, whereas the second one applies to all cycles between $1$ and $4$ that do not contain $H$, respectively. We are able to correctly identify that there is no hidden driving outside of $H$ and its connecting edges while also improving on the affinity estimator that does not discern whether $H$ has been visited or not. 

Finally we mention a case in which equation \eqref{eq:Delta_a} fails to detect hidden driving, \textit{i.e.}, $\Delta a=0$ despite $\aff\neq 0$. If the driven cycle is connected to the remaining network only through a single state as shown in the example of Figure~\ref{fig:networks}d), then its affinity does not result in a time-dependent $a(t)$. Put more generally, if a cycle can be located inside a compound state with direction-time independence \cite{wang07a, erte22} for all transitions to and from it, we are not able to infer its irreversibility from observables located entirely outside the compound state.


\section{Intricacies of odd variables}
\label{sec:justification}

The identification of entropy production is particularly subtle in the case of odd observables since the observations $I$ and $\rev{I}$ denote genuinely different objects. In the following, we examine two model classes based on such odd observables in detail.


\subsection{Underdamped Langevin dynamics}
\label{sec:underdamped}

\begin{figure}[t]
\begin{center}
    \includegraphics[scale=0.85]{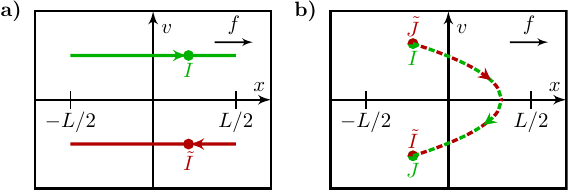}
    \caption{Sketches of underdamped trajectories in phase space. a) A trajectory $\Gamma_a$ that starts and ends in $I$ and completes the circle exactly once (green line)  and its time reverse (red line). b) A trajectory $\Gamma_b$ for which the velocity decreases linearly in time until the particle returns to its starting point (green curve). Its time reverse is identical to the forward trajectory (red curve) .}
    \label{fig:underdamped_traj}
\end{center}
\end{figure}	

For underdamped Langevin dynamics, the physically correct entropy production can be derived from the laws of stochastic energetics \cite{seki10}. However, a detailed fluctuation theorem of the form \eqref{eq:DFT} does not hold \cite{spin12, spin12a, lee12, fisc20}. 
In the following, we show and illustrate with some simple and well-understood examples that our approach is consistent with requirements for a physically meaningful entropy production in underdamped Langevin systems.

Consider an underdamped Brownian particle of mass $M$ on a one-dimensional ring subject to a potential $V$ and additionally to a non-conservative external force $f$ in a non-equilibrium steady state. It is described by the Langevin equation
\begin{align}
    M \partial _t^2 x = -\fric \partial_t x - \partial_x V(x) + f + M\xi(t)
,\end{align} in which $\xi$ models uncorrelated Gaussian white noise
\begin{align}
\mean{\xi(t)} & = 0 \\
\mean{\xi(t)\xi(t')} & = (2\fric T/M^2) \delta(t - t')
.\end{align}
In this set-up $\fric$ and $T$ denote the friction constant, which is not to be confused with the microscopic trajectories, and the temperature in units of energy, \textit{i.e.}, $k_B = 1$, respectively.

Its entropy production has two contributions, one due to the stochastic entropy production $\Delta S_{st}$ and one associated with dissipation into the medium $\Delta S_m$. First, we examine the entropy production along a trajectory $\Gamma_a$ that starts with $I = (x_0, v_0)$ at time $T=0$ and ends with $J = I$ after completing the cycle of length $L$ at time $T$ exactly once, as shown in Figure~\ref{fig:underdamped_traj}a) for $x_0 = -L/2 = L/2$. The stochastic contribution has to vanish since the initial and final states are identical. Therefore, we expect the total entropy production to be the cycle affinity $\mathcal{A}=fL$. Indeed, this is consistent with equation \eqref{eq:DeltaS_snip}, which here becomes
\begin{align} \label{eq:S_underdamped_1}
    \Delta S = \ln \frac{P(I)}{P(J)} + \ln\frac{\Path{\Gamma|I}}{\Path{\rev{\Gamma}|\rev{J}}} = \Delta S_{st} + \Delta S_{m}
.\end{align}
with $\Delta S_{st}=0$ and $\Delta S_{m}= \int_0^T f \dot{x} \d t = fL$. 

In contrast, the tentative identification 
\begin{align}
    \Delta S_{st}^{(i)} = \ln (P(I)/P(\rev{J}))
,\end{align}
for which $\Delta S^{(i)}\equiv\Delta S_m+ \Delta S_{st}^{(i)}$ would obey a detailed fluctuation theorem of the form \eqref{eq:DFT}, leads to $\Delta S_{st}^{(i)}\neq 0$ without a clear physical interpretation. First, this identification $\Delta S_{st}^{(i)}$ would not warrant the property of the stochastic entropy to purely depend on the state of the system. Second, the entropy production would not be additive. The sum of the entropy productions of two trajectories of the form shown in Figure~\ref{fig:underdamped_traj}a) would not be the same as the entropy production of the corresponding joint trajectory that completes the cycle twice.

A second identification of $\Delta S_{st}$ that reproduces the correct entropy production for the particular trajectory $\Gamma_a$ would be 
\begin{align}
    \Delta S_{st}^{(ii)} =  \ln (P(\rev{I})/P(\rev{J}))
.\end{align}
However, we can eliminate this possibility by considering a trajectory $\Gamma_b$ from $I = (x_0, v_0)$ to $J = (x_0, -v_0)$ where the velocity initially is parallel to the non-conservative driving force and then linearly decreases with time as shown in Figure~\ref{fig:underdamped_traj}b). This trajectory is identical to its time reverse and, therefore, satisfies $\Path{\Gamma|I} = \Path{\rev{\Gamma}|\rev{J}}$ by construction. Due to the negative velocity, the final state of this trajectory is less probable then the initial one, which implies that the stochastic entropy should increase, as is indeed the case for
\begin{equation} \label{eq:sto_underdamped_b}
    \Delta S_{st} = \ln \frac{P(I)}{P(J)} = \ln \frac{P(I)}{P(\rev{I})}
\end{equation}
while the alternative $\Delta S_{st}^{(ii)}$ gives the wrong sign. 

Finally, note that $-\ln (P(\rev{I})/P(\rev{J}))$ can be identified as the stochastic entropy production of the reverse trajectory in agreement with our formalism by applying equation~\eqref{eq:sto_underdamped_b} to $\rev{\traj}=\rev{J}\to\rev{I}$.



\subsection{Observed transitions}
\label{sec:transitions}
Transitions on a Markov network form a second class of systems with a description based on odd observables. We assume a system in a stationary state with only a few observable edges similar to the settings described in \cite{vdm22, haru22}. In this case, we can show that equation \eqref{eq:DeltaS_snip} is the correct identification of the entropy production through explicit calculations. We consider a coarse-grained trajectory $\traj=I\to J$ that starts with transition $I$ and ends with transition $J$. The corresponding microscopic trajectories share the form
\begin{equation}
    \gamma = i \overset{I}{\to} j \to \cdots \to k \overset{J}{\to} l
\end{equation}
where $i$, $j$ and $k$, $l$ are the Markov states associated with the transitions $I$ and $J$, respectively. These occur with rates
\begin{equation} \label{eq:app:trans_P}
P(I)=p_ik_{ij} \;\;\text{ and }\;\; P(J)=p_kk_{kl},
\end{equation}
where $p_i$ denotes the probability to find the system in microstate $i$. The remaining path weights read
\begin{equation} \label{eq:app:trans_Path1}
    \Path{I\to J|I}=\Path{j\to k|j}k_{kl} 
\end{equation}
and
\begin{equation} \label{eq:app:trans_Path2}
    \Path{\rev{J}\to \rev{I}|\rev{J}}=\Path{k\to j|k}k_{ji}.
\end{equation}
The term $\Path{j\to k|j}$ denotes the path weight for the trajectory starting immediately after entry into $j$ until immediately before exiting $k$, given that the system starts in $j$. Inserting equations \eqref{eq:app:trans_P}, \eqref{eq:app:trans_Path1} and \eqref{eq:app:trans_Path2} into the coarse-grained entropy production \eqref{eq:DeltaS_snip} yields
\begin{align*} \label{eq.app:trans_entropy}
    \Delta S[\traj] &= \ln\frac{P(I)\Path{I\to J|I}}{P(J)\Path{\rev{J}\to \rev{I}|\rev{J}}}
    = \ln\frac{p_ik_{ij}\Path{j\to k|j}k_{kl}}{p_kk_{kl}\Path{k\to j|k}k_{ji}} \\
    &= \ln\frac{p_ik_{ij}}{p_jk_{ji}}+\ln\frac{\Path{j\to k}}{\Path{k\to j}} \numberthis
,\end{align*}
where we identify $\Delta S[\traj]$ as the sum of the entropy production of the initial event $I$ and the entropy production of the remaining part of the trajectory until immediately before the concluding event $J$. The final event is not included in the entropy production $\Delta S[\traj]$ in accordance with Section~\ref{sec:CG_entropy}.

By analogy to the cases of Markov networks and underdamped Langevin dynamics, equation \eqref{eq.app:trans_entropy} suggests an identification of stochastic and medium entropy production in the form
\begin{align}
    \Delta S = \ln\frac{P(I)}{P(J)} + \ln \frac{\Path{I\to J|I}}{\Path{\rev{J}\to \rev{I}|\rev{J}}}= \Delta S_{st} + \Delta S_{m}
.\end{align}
This identification of $\Delta S_{st}$ and $\Delta S_{m}$ must be clearly distinguished from the identification $\Delta s = \Delta s_{st} + \Delta s_{m}$ that can be made on the microscopic level for the Markov network \cite{seif05a}. We emphasize that the two identifications do not coincide even if the coarse graining retains the full entropy production, \textit{i.e.}, even if $\mean{\Delta s} = \mean{\Delta S}$, which is the case for, \textit{e.g.}, unicyclic Markov networks with a single observed transition \cite{vdm22}. In such a scenario, we have two physically sensible splittings into stochastic and medium entropy production. Thus, we interpret that the conception of 'system' and 'medium' can depend on the available information even if both descriptions apply to the same physical process on the microscopic level. For example, while stochastic entropy on the microscopic level of the Markov network may change whenever the actual state $i$ of the system changes, a corresponding quantity on the coarse-grained level updates when an observed transition $I$ is registered. 


\subsection{Discussion}
\label{sec:discussion}

The above analysis of odd variables shows that even when considering stationary situations only, entropy production, in general, does not obey the detailed fluctuation theorem that applies to a Markovian dynamics \cite{seif05a}. Even if such an underlying Markovian description exists and equation \eqref{eq:DeltaS_snip} can be applied, we cannot expect a detailed fluctuation theorem of the same functional form as on the microscopic level, because the coarse-grained level involves different observables. However, validity is granted if all observables are even under time reversal, as seen by plugging $P(\rev{\mev{J}})=P(\mev{J})$ into equation \eqref{eq:DeltaS_snip}. Thus, a refined understanding of the detailed fluctuation theorem may be as a symmmetry associated with the special case of a description based on even observables in a Markovian dynamics rather than a general property of entropy production itself. 

For underdamped Langevin dynamics in particular, entropy production is often expressed using some 'modified' path weight for the reverse process \cite{spin12, spin12a, fisc20}, which leads to a formulation that appears similar to the detailed fluctuation theorem. However, this modified path weight and, therefore, the entire approach lack a clear physical interpretation. In light of the above insights regarding the detailed fluctuation theorem, it seems neither useful nor instructive to employ such a 'modified' path weight.

An immediate consequence is that an expression of the form $\ln\left(\Path{\gamma}/\Path{\rev{\gamma}}\right)$ or $\ln(\Path{\traj}/\Path{\rev{\traj}})$ cannot be regarded as a guiding principle for the identification of entropy production in arbitrary systems. For any system that on the underlying level fulfills the conditions described in Section~\ref{sec:micro}, one should resort to the identification in equation~\eqref{eq:DeltaS_snip}. Beyond these, the identification of an entropy production requires knowledge of the underlying energetics and, additionally, considerations similar to the ones in Section~\ref{sec:underdamped}. The analysis of scenarios for which we understand the corresponding entropy production physically, such as in equilibrium or for simple trajectories, gives rise to consistency conditions that entropy production has to fulfill. 

\section{Concluding perspective}
\label{sec:conclusion}


In this work, we have established requirements to identify a fluctuating entropy production on a coarse-grained level. This concept offers practical advantages in resolving sources of irreversibility. Depending on the information available, we can quantify the contribution of particular snippets of a long trajectory to the total entropy production and localize hidden driving with the aid of bounds on their affinity. As our identification is model-free, we put particular emphasis on the subtle case of odd variables.

Our framework does not rely on distinguishing slow and fast degrees of freedom. Such knowledge typically is hard to come by, particularly in the realistic case where some degrees of freedom are hidden. Thus, it is not necessary to identify  a 'correct fundamental layer', \textit{i.e.}, a system comprising all slow-moving degrees of freedom which is surrounded by an environment that equilibrates on a faster timescale. Instead, the hidden parts of the system and the surrounding system are treated in the same way without making particular distinctions beforehand.

Finally, we point out that, in the general case, one should not expect to identify a physically meaningful entropy production on the coarse-grained level solely based on our approach in the absence of Markovian events, since our formalism necessarily requires a coarse-grained trajectory to start and end with a Markovian event. Applying equation \eqref{eq:DeltaS_snip} to arbitrary sections of a coarse-grained trajectory would generally imply a violation of the properties mentioned in Section~\ref{sec:CG_entropy}, which would undermine the corresponding physical interpretation. In particular, the correspondence between microscopic and coarse-grained entropy production relies on the defining property of Markovian events.



The results of this paper allow for future research beyond the insights already demonstrated. As knowledge about the dynamics of the hidden parts of the system is not required, the improved flexibility of our framework should prove useful for applications across different model classes. Nevertheless, incorporating additional knowledge remains possible, as demonstrated for the estimation of cycle affinities when supposing an underlying network of Markov states. In a realistic biochemical set-up, cycle affinities might be constrained to integer multiples of, say, the free energy released in hydrolysis of one ATP. Thus, if some microscopic states or transitions can be observed directly, the established affinity bounds that are able to distinguish different pathways provide a qualitative tool to localize the ATP-driven cycles in the biochemical network.

We expect that the concept of spatially and temporally localized entropy estimation can be combined successfully with other related techniques. 
Inferring entropy production through waiting times in snippets provides a physical interpretation for the waiting time distributions that 
have received attention in stochastic thermodynamics in different contexts. 
For example, recently discovered thermodynamic bounds on waiting time distributions and correlation functions \cite{ohga23, kolc23, dech23, dech23a, shir23} might be combined with the notion of fluctuating entropy production to infer not only averages but full distributions of thermodynamic quantities.

In addition, our framework applies to systems in which the state of the system is known at some times, but the full dynamics remains inaccessible or unknown, which generalizes the paradigmatic cases of fully accessible Markov networks and overdamped Langevin equations. With this new perspective, we might speculate whether other concepts of stochastic thermodynamics beyond total entropy production like an identification of intrinsic and medium entropy production or stochastic energetics can also be extended to genuinely coarse-grained descriptions.

\appendix

\onecolumngrid



\FloatBarrier


\section{Proof of the equality \eqref{eq:s_exp_mean}}
\label{sec:proof_exp_mean}
We consider a coarse-grained trajectory $\traj$, which starts with $\mev{I}$ and ends with $\mev{J}$. Inserting the definition of the microscopic entropy production \eqref{eq:micro_delta_s} into the mean on the right hand side of equation \eqref{eq:s_exp_mean} yields
\begin{equation}
    \mean{e^{-\Delta s}|\traj}=\sum_{\gamma\in\traj}\Path{\gamma|\traj}\frac{P(\gamma_1)\Path{\rev{\gamma}|\rev{\gamma_1}}}{P(\gamma_0)\Path{\gamma|\gamma_0}}=\sum_{\gamma\in\traj}\Path{\gamma|\traj}\frac{P(\mev{J})\Path{\rev{\gamma}|\rev{\mev{J}}}}{P(\mev{I})\Path{\gamma|\mev{I}}}.
\end{equation}
For the second equality we used that $\gamma$ is initialized by the Markovian event $\mev{I}$, \textit{i.e.}, $\gamma_0=\mev{I}$, and, likewise, $\gamma_1=\mev{J}$. Since the coarse graining defines a unique mapping $\gamma\mapsto\traj$, each microscopic trajectory $\gamma$ implies its corresponding coarse-grained trajectory $\Gamma$, which allows to calculate
\begin{equation}
    \sum_{\gamma\in\traj}\Path{\gamma|\traj}\frac{P(\mev{J})\Path{\rev{\gamma}|\rev{\mev{J}}}}{P(\mev{I})\Path{\gamma|\mev{I}}} 
    = \sum_{\gamma\in\traj}\frac{\Path{\gamma}}{\Path{\traj}}\frac{P(\mev{J})\Path{\rev{\gamma}|\rev{\mev{J}}}}{\Path{\gamma}} 
    = \frac{P(\mev{J})}{P(\mev{I})\Path{\traj|\mev{I}}}\sum_{\gamma\in\traj}\Path{\rev{\gamma}|\rev{\mev{J}}}
    = \frac{P(\mev{J})\Path{\rev{\traj}|\rev{\mev{J}}}}{P(\mev{I})\Path{\traj|\mev{I}}}
    = e^{-\Delta S[\traj]}.
\end{equation}


\section{Proof of the affinity bound \eqref{eq:affinity bound}}
\label{sec:affinity_proof}
For the proof of the bound \eqref{eq:affinity bound}, we assume an underlying Markov network in a stationary state, some coarse graining $\cg$ and a time-reversal operation $\tr$ with $\tr\gamma=\rev{\gamma}$, which is an involution. We consider a second involution $\R$, which has the property that it is not distinguishable from the time reversal $\tr$ under coarse graining, \textit{i.e.},
\begin{equation} \label{eq:app:R}
    \begin{tikzcd}
        \gamma \arrow[r, "\tr"] \arrow[d, "\R"] & \rev{\gamma} \arrow[d, "\cg"] \\
        \mathcal{R} \gamma \arrow[r, "\cg"] & \rev{\traj}
    \end{tikzcd}
.\end{equation}
Since additionally $\R$ is a bijection, we conclude
\begin{equation} \label{eq:app:P1}
    \Path{\rev{\traj}|\rev{J}} = \sum_{\gamma \in \traj} \Path{\rev{\gamma}|\rev{J}} = \sum_{\gamma \in \traj} \Path{\mathcal{R} \gamma|\widetilde{J}}.
\end{equation}
For a trajectory $\gamma$ that starts with the Markovian event $I$ and ends with the Markovian event $J$, we define 
\begin{equation} \label{eq:app:A}
    \A[\gamma] \equiv \ln\frac{\Path{\gamma|I}}{\Path{\R\gamma|\rev{J}}}.
\end{equation}
Note that $\rev{J}$ necessarily is the initial event of $\R\gamma$ due to the relation \eqref{eq:app:R}. Using the definition \eqref{eq:app:A}, we rewrite
\begin{equation} \label{eq:app:P2}
        \Path{\traj|I} = \sum_{\gamma \in \traj} \Path{\gamma|I} = \sum_{\gamma \in \traj} \Path{\mathcal{R} \gamma|\widetilde{J}} e^{\A[\gamma]}.
\end{equation}
Combining equations \eqref{eq:app:P1} and \eqref{eq:app:P2}, we find 
\begin{equation} \label{eq:app:P3}
    \frac{\Path{\traj|I} }{\Path{\rev{\traj}|\rev{J}}} = \frac{\sum_{\gamma \in \traj} \Path{\mathcal{R} \gamma|\widetilde{J}} e^{\A[\gamma]}}{\sum_{\gamma \in \traj} \Path{\mathcal{R} \gamma|\widetilde{J}}} = \mean{e^{\A[\gamma]}}_\text{aux},
\end{equation}
where we identify the expression in the middle as a mean with respect to some auxiliary probability measure. This relation implies
\begin{equation} \label{eq:app:P4}
    \inf_{\gamma\in\traj} e^{\Delta \A[\gamma]} \leq \mean{e^{\Delta \A[\gamma]}} \leq \sup_{\gamma\in\traj} e^{\Delta \A[\gamma]}.
\end{equation}
Inserting equation \eqref{eq:app:P3} into \eqref{eq:app:P4} yields
\begin{equation} \label{eq:app:P5}
    \inf_{\gamma\in\traj} \A[\gamma] \leq \ln\frac{\Path{\traj|I}}{\Path{\rev{\traj}|\rev{J}}} \leq \sup_{\gamma\in\traj} \A[\gamma]
,\end{equation}
where we also use the monotonicity of the logarithm. 
The inequalities \eqref{eq:app:P5} hold for any coarse-grained trajectory $\traj$. By considering all trajectories that start in $I$ and end in $J$, denoted symbolically as $\gamma|I\to J$, we obtain
\begin{equation}
    \sup_{\gamma|I\to J} \A[\gamma] - \inf_{\gamma|I\to J} \A[\gamma] \geq \sup_{\traj|I\to J} \ln \frac{\Path{\traj|I} }{\Path{\rev{\traj}|\rev{J}}} - \inf_{\traj|I\to J} \ln \frac{\Path{\traj|I}}{\Path{\rev{\Gamma}|\rev{J}}}.
\end{equation}
Since the initial and final events are fixed, the probabilities $P(I)$ and $P(J)$ are identical for all these trajectories, so that the duration of the snippet becomes the only remaining variable, which finally results in
\begin{equation} \label{eq:app:bound_intermediate}
    \sup_{\gamma|I\to J} \A[\gamma] - \inf_{\gamma|I\to J} \A[\gamma] \geq \sup_t\ln \frac{P(I)\Path{\traj|I} }{P(J)\Path{\rev{\traj}|\rev{J}}} - \inf_t \ln \frac{P(I)\Path{\traj|I}}{P(J)\Path{\rev{\Gamma}|\rev{J}}}=\sup_t\Delta S[I\xrightarrow{t}J] - \inf_t \Delta S[I\xrightarrow{t}J].
\end{equation}

To extract a physically meaningful bound from the inequality \eqref{eq:app:bound_intermediate}, we need to explicitly construct $\R$, where we follow an algorithm similar to \cite{vdm22}. Consider the microscopic trajectory on the Markov network $\gamma=I\to k\to\cdots\to l\to J=I k\cdots l J$, where $I$ and $J$ are the Markovian events, in this case states or transitions. To contruct $\R\gamma$, we perform the following steps:
\begin{enumerate}[ {}1{)} ]
    \item Separate the trajectory into closed loops and the remaining parts. Starting after $I$ and stopping before $J$, if a state occurs more than once, we identify a closed loop as the intermediate section between the first and last appearance including the states at the beginning and end of the loop, 
    \begin{equation}
        (\cdots a  x b \cdots c x d\cdots)\mapsto (\cdots a)  (x b \cdots c x)  (d\cdots)
    \end{equation}
    It is not necessary to identify loops within loops.
    \item Reverse the order of sections, i.e.,
    \begin{equation}
        (a\cdots b)(cd\cdots ec)(f\cdots g) \mapsto (f\cdots g)(cd\cdots ec)(a\cdots b).
    \end{equation}
    \item Reverse the order of states within the sections that do not form a closed loop. These are the sections with differing initial and final states, i.e.,
    \begin{equation}
        (f\cdots g)(cd\cdots ec)(a\cdots b) \mapsto (g\cdots f)(cd\cdots ec)(b\cdots a).
    \end{equation}
    \item Reverse $I$ and $J$, i.e.,
    \begin{equation}
        I\mapsto\rev{I}, J\mapsto\rev{J}.
    \end{equation}
    \item Merge the resulting section.
\end{enumerate}
Note that the residence times in each state are not affected by $\R$.

We define $\gamma^\text{trim}$ as the trimmed trajectory that results from removing the closed loops of $\gamma$ in the form of
\begin{equation}
    \gamma=abcdecfg \implies \gamma^\text{trim} = abcfg,
\end{equation}
where the loops are identified in the same way as in step 1).
With this explicit definition of $\R$, we identify
\begin{equation}
    \A[\gamma] = \aff[\gamma^\text{trim}] + C(I,J)
\end{equation}
with the affinity
\begin{equation}
    \aff[\gamma] \equiv \sum_{(ij)\in\gamma} \ln\frac{k_{ij}}{k_{ji}}
\end{equation}
where the sum runs over all transitions in $\gamma$. In the general case, a boundary term $C(I,J)$ may be present, which exclusively depends on $I$ and $J$ and therefore is irrelevant after taking differences in \eqref{eq:app:bound_intermediate}. Note that we can identify this difference as an extremal cycle affinity, \textit{i.e.},
\begin{equation} \label{eq:app:ident_aff}
    \sup_\gamma A[\gamma] - \inf_\gamma A[\gamma] = \max_\cyc\mathcal{A_\cyc}
,\end{equation}
where the maximum runs over all cycles that can be written as $\gamma^\text{trim}_1 \to \tr\gamma_2^\text{trim} = \gamma^\text{trim}_1\to\rev{\gamma_2^\text{trim}}$. Note that each affinity $\mathcal{A_\cyc}$ occurs with either sign, so that the maximum selects the cycle affinity with the highest absolute value. By inserting equation \eqref{eq:app:ident_aff} into the bound \eqref{eq:app:bound_intermediate}, we finally arrive at
\begin{equation}
    \max_\cyc\mathcal{A_\cyc} \geq \sup_t\Delta S[I\xrightarrow{t}J] - \inf_t \Delta S[I\xrightarrow{t}J],
\end{equation}
which is equation \eqref{eq:affinity bound} in the main text.

\section{Rates}
\label{sec:rates}

The rates for the network shown in Figure~\ref{fig:networks}a) are $k_{12}=0.6$, $k_{21} = 2$,  $k_{23} = 2.1$, $k_{32} = 5$, $k_{26} = 1.4$, $k_{62} = 2$, $k_{34} = 0.3$, $k_{43} = 3$, $k_{35} = 0.9$, $k_{53} = 4.3$, $k_{36} = 1$, $k_{63} = 1.6$, $k_{45} = 0.5$, $k_{54} = 2$, $k_{56} = 0.8$, $k_{65} = 4$, $k_{61} = 1.8$ and $k_{16} = 3$.

The rates for the network shown in Figure~\ref{fig:networks}c) are $k_{12}=1$, $k_{15}=1.5$, $k_{21}=1$, $k_{23}=1$, $k_{26}=0.5$, $k_{32}=1$, $k_{34}=1$, $k_{37}=0.5$, $k_{43}=1$, $k_{47}=0.5$, $k_{51}=1.5$, $k_{56}=0.5$, $k_{58}=0.5$, $k_{62}=0.5$, $k_{65}=0.5$, $k_{67}=0.5$, $k_{68}=0.1$, $k_{69}=0.5$, $k_{73}=0.5$, $k_{74}=0.5$, $k_{76}=0.5$, $k_{79}=0.1$, $k_{85}=0.1$, $k_{86}=0.5$, $k_{89}=3$, $k_{96}=0.1$, $k_{97}=2$ and $k_{98}=4$.

The rates for the network shown in Figure~\ref{fig:aff_chem}b) are $k_{12}=2.5$, $k_{21}=2.5$, $k_{23}=0.7$, $k_{25}=0.1$, $k_{32}=0.3$, $k_{34}=0.7$, $k_{43}=0.3$, $k_{45}=0.7$, $k_{52}=1$, $k_{54}=0.3$, $k_{56}=2.5$ and $k_{65}=2.5$

\newpage

\twocolumngrid


%

\end{document}